\documentclass[]{emulateapj}

\usepackage{natbib}
\usepackage{amsmath}
\usepackage{txfonts}
\usepackage{gensymb}
\usepackage{mathrsfs}
\usepackage{epstopdf}
\usepackage{txfonts}
\usepackage{graphicx,color}

\newcommand{\rH}{r_{\rm H}}

\newcommand{\rs}{r_{\rm s}}
\newcommand{\cs}{c_{\rm s}}

\newcommand{\msunyr}{M_{\odot}~\rm yr^{-1}}

\newcommand{\eqnref}[1]{Equation \ref{#1}}

\newcommand{\figref}[1]{Figure \ref{#1}}
\newcommand{\tabref}[1]{Table \ref{#1}}

\shorttitle{Save the Planet, Feed the Star}
\shortauthors{Fung \& Chiang}

\begin{document}
\title{Save the Planet, Feed the Star: \\ How Super-Earths Survive Migration and Drive Disk Accretion}
\author{Jeffrey Fung\altaffilmark{1} \& Eugene Chiang}
\affil{Department of Astronomy, University of California at Berkeley, Campbell Hall, Berkeley, CA 94720-3411}
\altaffiltext{1}{NASA Sagan Fellow}

\email{email: jeffrey.fung@berkeley.edu}

\begin{abstract}

Two longstanding problems in planet formation
include (1) understanding how planets survive migration,
and (2) articulating the process
by which protoplanetary disks disperse---and in particular
how they accrete onto their central stars.
We can go a long way toward solving both problems 
if the disk gas surrounding planets
has no intrinsic diffusivity (``viscosity'').
In inviscid, laminar disks, a planet readily repels gas
away from its orbit.
On short timescales,
zero viscosity gas accumulates
inside a planet's orbit to slow Type I migration by orders 
of magnitude. On longer timescales,
multiple super-Earths (distributed between,
say, $\sim$0.1--10 AU) can torque 
inviscid gas out of interplanetary space, either inward 
to feed their stars, or outward to be blown away in a wind.
We explore this picture with 2D hydrodynamics simulations
of Earths and super-Earths embedded in inviscid disks,
confirming their slow/stalled migration even under
gas-rich conditions,
and showing that disk transport rates
range up
to $\sim$$10^{-7} \msunyr$ and scale as 
$\dot{M} \propto \Sigma M_{\rm p}^{3/2}$, 
where $\Sigma$ is the disk surface density 
and $M_{\rm p}$ is the planet mass.
Gas initially sandwiched between two
planets is torqued past both into the inner
and outer disks.
In sum,
sufficiently compact systems of super-Earths can
clear their natal disk gas, 
in a dispersal history
that may be complicated and non-steady, but which conceivably
leads over Myr timescales to large gas depletions 
similar to those characterizing transition disks.
\end{abstract}


\section{Introduction}
\label{sec:intro}

Protoplanetary disks have two jobs: make planets and 
feed their host stars. The first task is frustrated
by migration: disk torques force 
planetary orbits to decay \citep[e.g.,][]{Kley12},
evacuating the very regions where planets
are observed in abundance 
\citep[$\sim$0.1--10 AU from the star; e.g.,][]
{Clanton14,Burke15,Christiansen15,Dressing15}.
Fulfilling the second task requires a mechanism
to transport away the disk's angular momentum.
Magnetic torques are promising but depend
on seed fields of uncertain provenance \citep[e.g.,][]{Bai16}.

\citet{Goodman01} proposed that the two problems are actually
one: that planets themselves---if they can survive migration---can 
provide an effective source
of disk viscosity by exciting density waves that transport
angular momentum outward. \citet{Sari04} emphasized that
such planets must be massive enough to open gaps and
avoid Type I migration.
They focused on giant Jupiter-mass planets, a demographic
that is now understood to be rare \citep[e.g.,][]{Cumming08}.
In this paper, we turn our attention to super-Earths: bodies
of mass 1--10 $M_\oplus$ that have been
discovered by {\it Kepler} to be
relatively commonplace \citep[e.g.,][]{Fressin13}.

Can super-Earths avoid Type I migration? Yes---if
their disks are sufficiently inviscid. The
dependence of the Type I drift rate on disk viscosity
is perhaps under-appreciated, as it is not explicitly
called out in the typically quoted Type I formula
(see, e.g., \citealt{Kley12}).
Crucially, without an intrinsic disk viscosity to 
smooth away the planet's perturbations
to the disk's surface density,
a pile-up of disk material ahead of a migrating planet
exerts a ``feedback'' torque that
slows and can even stall migration
\citep[][]{Hourigan84,Ward89}.
\citet[][see his equation 53]{rafikov02b} calculated
that for inviscid
disks in which planet-driven waves dissipate 
by steepening into shocks, the critical
planet mass above which Type I migration shuts off is:
\begin{equation}
\label{eqn:m_cr}
M_{\rm cr} \simeq
4 \left(\frac{h_{\rm p}/r_{\rm p}}{0.035}\right)^{3}
\left(\frac{M_\ast}{M_{\odot}}\right)
\left(\frac{ \Sigma_{\rm p} r_{\rm p}^2 / M_\ast}{10^{-3}}\right)^{\frac{5}{13}} M_\oplus \, ,
\end{equation}
where $r$ and $h$ are the disk radius and scale height,
$\Sigma$ is the disk gas surface density, $M_\ast$ is the
central stellar mass, and the subscript ${\rm p}$
indicates evaluation near the planet's position. 
\citet{Li09} and \citet{Yu10} have performed numerical
simulations supporting the analytic calculations by 
\citet{rafikov02b}, and confirming that super-Earths
in low-viscosity disks migrate much more slowly (and erratically)
than is predicted by Type I.

A planet of mass $M_{\rm p}$
at $r_{\rm p}$ drives a disk
mass transport
rate $\dot{M}$ at
distance $r$ of
\begin{equation}
\label{eqn:mdot}
\dot{M}(r) = -\frac{2F_0}{l} r \frac{\partial \varphi(r)}{\partial r} 
\end{equation}
where
\begin{equation}
\label{eqn:torque}
F_0 = \Sigma_{\rm p} r_{\rm p}^2 l_{\rm p} \Omega_{\rm p}  \left(\frac{M_{\rm p}}{M_*}\right)^2 \left(\frac{h_{\rm p}}{r_{\rm p}}\right)^{-3} 
\end{equation}
measures the total angular momentum carried away per time
by planet-driven waves (a.k.a.~the total integrated
one-sided Lindblad torque),
$\Omega$ is the orbital frequency, and
$l = \Omega r^2$ is the specific angular momentum.
The dimensionless function $\varphi(r)$ 
describes how waves, as they travel away from the planet,
damp with distance, depositing their
angular momentum to disk gas 
and thereby propelling material
radially.
From \citet[][see his equations B1 and 33]{rafikov02b},
\begin{equation}
\label{eqn:phi}
\varphi \sim \left( \frac{M_{\rm p}}{M_{\rm thermal}} \right)^{-\frac{1}{2}} \left( \frac{|r_{\rm p} - r|}{h_{\rm p}} \right)^{-\frac{5}{4}} \,,
\end{equation}
valid for $M_{\rm p} \lesssim M_{\rm thermal}
\equiv (h_{\rm p}/r_{\rm p})^{3} M_\ast$ and
$\varphi \lesssim 1$ (i.e., distances far enough from the planet
that the waves are dissipating in weak shocks), and where
we have ignored order-unity constants and
all radial variations in $h$, $\Sigma$, and gas sound speed.
It follows that
\begin{equation}
\label{eqn:ana_mdot}
\dot{M}(r) \sim {\rm sign}(r-r_{\rm p})~\Sigma_{\rm p} r_{\rm p}^2 \Omega_{\rm p} \left(\frac{M_{\rm p}}{M_{\ast}}\right)^{\frac{3}{2}}
\left(\frac{h_{\rm p}}{r_{\rm p}}\right)^{-\frac{5}{2}}
\left(\frac{|r_{\rm p}-r|}{h_{\rm p}}\right)^{-\frac{9}{4}} \, .
\end{equation}
At $r < r_{\rm p}$, $\dot{M}$ as given by
Equation (\ref{eqn:ana_mdot}) is negative
(mass flows inward), and vice versa; a planet
tends to repel material away from itself.
To avoid minus signs, we will ignore this formal
sign convention so that all our reported values for $\dot{M}$
will be positive and understood to be inward unless
otherwise indicated.

Note how $\dot{M} \propto M_{\rm p}^{3/2}$ and not $M_{\rm p}^2$.
Although the total Lindblad
torque scales as $M_{\rm p}^2$ (Equation \ref{eqn:torque}),
that torque is distributed over a distance that increases
with decreasing $M_{\rm p}$ (as $M_{\rm p}^{-2/5}$, as can be
seen by solving for $|r_{\rm p} - r|$ in terms of $M_{\rm p}$
at fixed $\varphi$ in Equation \ref{eqn:phi}). 
Thus at fixed distance away from the planet, $\dot{M}$ increases
with $M_{\rm p}$ with a power less than 2.

Inserting $M_{\rm p}  = 10M_\oplus$
and other nominal parameters (for $r < r_{\rm p}$) into
(\ref{eqn:ana_mdot}) yields
\begin{align}
\label{eqn:ana_mdot_wnum}
\nonumber
\dot{M}(r) \sim 10^{-8} &\left(\frac{\Sigma_{\rm p} r_{\rm p}^2}{10^{-3} M_\odot}\right) \left(\frac{2\pi/\Omega_{\rm p}}{1\,{\rm yr}}\right)^{-1} \left(\frac{M_{\rm p}/M_*}{3 \times 10^{-5}}\right)^{\frac{3}{2}} \left(\frac{h_{\rm p}/r_{\rm p}}{0.035}\right)^{-\frac{1}{4}} \\
& \left(\frac{(r_{\rm p} - r)/r_{\rm p}}{0.5}\right)^{-\frac{9}{4}} \msunyr \, ,
\end{align}
comparable to accretion rates measured for classical T Tauri stars
\citep[e.g.,][]{Calvet05,Hartmann06,Sicilia10,Ingleby13}.
Note how weakly $\dot{M}$ depends on $h_{\rm p}/r_{\rm p}$,
underscoring how $\dot{M}$ does not scale simply
as the total Lindblad torque in \eqnref{eqn:torque} 
(which scales as $(h_{\rm p}/r_{\rm p})^{-3}$),
but depends also on the distance over which that torque
is exerted, as we have described above.

The above considerations indicate that with super-Earths
we might have our cake (survive migration) and eat it, 
too (drive disk accretion). Of course, a single 
super-Earth is insufficient because its reach is too 
short ($\dot{M}$ drops as $|r-r_{\rm p}|^{-9/4}$).
Multiple super-Earths are needed to shuttle the
accretion flow from distances of a few AU
down to the stellar radius. Reality will be non-steady
and likely messy (see, e.g., Figure 4 of
\citealt{Rafikov02a}), with material between
adjacent planets having a fate that is not obvious:
does the sandwiched gas drain inward,
or does the inner planet hold back material
pushed inward by the outer planet?
And to what extent do super-Earths migrate with
the accretion flow they drive?

Here we explore these questions
using fully non-linear, 2D
hydrodynamical simulations of super-Earths embedded
in inviscid disks. We measure the migration
histories $r_{\rm p}(t)$ and accretion rates $\dot{M}$
in simulations containing 1 or 2 super-Earths,
experimenting with varying the disk surface density and
the planet mass to test \eqnref{eqn:ana_mdot_wnum}.
Our numerical methods are given in Section \ref{sec:numerics}.
Results are presented in Section \ref{sec:results} and
placed into broader context in Section \ref{sec:conclude}.

\section{Numerical Method}
\label{sec:numerics}
We use the graphics processing unit (GPU) accelerated
hydrodynamics code \texttt{PEnGUIn} \citep{MyThesis} to perform
2D simulations of disk-planet interactions.
It is a Lagrangian-remap shock-capturing code that uses the
piecewise parabolic method \citep{PPM} to solve the
continuity and momentum equations:
\begin{align}
\label{eqn:cont_eqn}
\frac{{\rm D}\Sigma}{{\rm D}t} &= -\Sigma\left(\nabla\cdot\mathbf{v}\right) \,,\\
\label{eqn:moment_eqn}
\frac{{\rm D}\mathbf{v}}{{\rm D}t} &= -\frac{1}{\Sigma}\nabla p + \frac{1}{\Sigma}\nabla\cdot\mathbb{T} - \nabla \Phi \,,
\end{align}
where $\Sigma$ is the gas surface density, $\mathbf{v}$ the
velocity field, $p$ the vertically averaged gas pressure,
$\mathbb{T}$ the Newtonian stress tensor, and $\Phi$ the
combined gravitational potential of the star and the planet(s).
We use a globally isothermal equation of state:
$p=\cs^2\Sigma$ with a spatially constant sound speed 
$\cs=0.035 v_{\rm K, 1\,AU} \simeq 1~\rm km~s^{-1}$
where $v_{\rm K, 1\,AU}$ is the Keplerian velocity
at 1 AU around a 1 $M_\odot$ star. (This $\cs$ corresponds
to a disk temperature of 300 K assuming a mean molecular weight
of 2.34.)

In a polar coordinate system (radius $r$, azimuth $\phi$)
centered on the star,
\begin{align}\label{eqn:potential}
\Phi &= -\frac{GM_*}{r} + \sum_{i=1}^{N_{\rm p}}\Phi_{{\rm p},i}  \\
\Phi_{{\rm p},i} &= -\frac{GM_{{\rm p}, i}}{\sqrt{r^2 + r_{{\rm p}, i}^2 - 2rr_{{\rm p}, i}\cos{\phi_{i}'} + r_{{\rm s}, i}^2}} + \frac{GM_{{\rm p}, i}~r\cos{\phi_{i}'}}{r_{{\rm p}, i}^2}
\end{align}
where $G$ is the gravitational constant,
$M_\ast = 1 M_\odot$ is the stellar mass,
the subscript $i$ labels each planet,
$N_{\rm p}$ is the
total number of planets, 
$M_{\rm p}$ is the planet mass, 
$\Phi_{{\rm p}}$ the planet's gravitational potential,
$r_{\rm p}$ the planet's radial coordinate, 
$\rs$ the smoothing length
of the planet's potential, and $\phi' = \phi-\phi_{\rm p}$ the
azimuthal separation from the planet.
The stress tensor $\mathbb{T}$ is proportional to the
kinematic viscosity $\nu$. Most of our simulations are of 
inviscid disks with $\nu=0$. For our viscous disk
simulations, we use $\nu=\alpha\cs h$, where the Shakura-Sunyaev
parameter $\alpha = 0.001$, $h = \cs/\Omega_{\rm K}$ is the
local scale height, and $\Omega_{\rm K} = \sqrt{GM_\ast/r^3}$
is the Keplerian orbital angular
frequency. At $r = 1$ AU, $h/r = 0.035$.
We set $\rs = 0.5 h$,
as is appropriate for 2D simulations \citep{Muller12}.

A given planet feels the gravitational force from the star,
the disk, and other planets.
The disk force on the planet is calculated by direct summation
over all mass
elements in the disk, with the ``background''
axisymmetric component of
the disk surface density subtracted off. Because the disk does not
feel its own gravity at all
(i.e., we ignore disk self-gravity;
see equation \ref{eqn:potential}), eliminating
this axisymmetric component in the disk-planet forcing
improves consistency between the motions
of the planets and the disk. Planet migration should be minimally
affected by this procedure, since the background
component of $\Sigma$ exerts
no torque. 
Spurious forces arising from within the planet's Hill
sphere
are
sometimes a concern if
this
region is under-resolved.
The Hill radius, $\rH = (M_{\rm p}/3M_*)^{1/3}$, 
ranges from 0.3 to
0.6 $h_{\rm p}$, similar to the smoothing length $\rs$.
We have
verified that the
torque generated 
within a radius of
0.5 $\rH$ from the planet is negligible, and so
we do not excise
the Hill sphere
in force calculations.
The planets' motions are integrated using a
kick-drift-kick leapfrog scheme, with the drift step occurring
synchronously with the hydrodynamics step; i.e., the planets'
positions are linear in time within a hydrodynamics step.

\vspace{0.5in}

\subsection{Initial and boundary conditions, and grid parameters}
\label{sec:initial}

\begin{deluxetable*}{ccccccccc}
\tablecaption{\label{tab:setup} Model Parameters}
\tablehead{Model \#&$M_{\rm p}~(M_{\oplus})$&$r_{\rm p,1}$ (AU)&$r_{\rm p,2}$ (AU)&$\Sigma_0$ (g $\rm cm^{-2}$) & $\alpha$ & $r_{\rm in}$ (AU)& $t_{\rm end}$ (years)}
\startdata 
1 & 10 & 1    & --   & $8.5\times 10^{3}$ & $10^{-3}$ & 0.4 & 700\\
2 & 10 & 1    & --   & $8.5\times 10^{3}$ & 0         & 0.4 & 5000\\
3 & 10 & 1    & 1.2  & $8.5\times 10^{3}$ & 0         & 0.4 & 5000\\
4 & 10 & 0.75 & 1.05 & 8.5                & 0         & 0.3 & 2000\\
5 & 3  & 0.75 & 1.05 & 8.5                & 0         & 0.3 & 2000\\
6 & 1  & 0.75 & 1.05 & 8.5                & 0         & 0.3 & 2000
\enddata
\tablecomments{$t_{\rm end}$ is the end time of a simulation, in units where the Keplerian orbital period at 1 AU is 1 year. Also, $r_{\rm p,1}$ and $r_{\rm p,2}$ are merely the initial planet locations at $t=0$; the planets are completely free to migrate in the simulations.}
\end{deluxetable*}

\tabref{tab:setup} lists the parameters used by our
6 models.
The disk is initialized with a power-law surface density:
\begin{equation}\label{eqn:initial_sigma}
\Sigma =  \Sigma_0 \left(\frac{r}{\rm AU} \right)^{-\frac{3}{2}}\, .
\end{equation}
We consider both gas-rich disks having
$\Sigma_0=8.5\times10^3~\rm g~cm^{-2}$
resembling the minimum-mass
extrasolar nebula (\citealt{Chiang13}),
and gas-poor disks having a surface density
1000$\times$ lower. The initial velocity field
is axisymmetric and Keplerian,
with corrections from gas pressure:
\begin{equation}\label{eqn:omega}
\Omega = \sqrt{\Omega_{\rm K}^2 + \frac{1}{r\Sigma}\frac{{\rm d} p}{{\rm d} r}} \,.
\end{equation}
One planet, whose mass is increased gradually over the
first 10 yr of the simulation
to the full value of $M_{\rm p}$ (either
1, 3, or 10 $M_\oplus$), 
is placed initially at
$r = r_{\rm p, 1}$ (either
1 or 0.75 AU) 
and $\phi_{\rm p, 1} = \pi$.
In two-planet models, we place a second planet
of equal mass to the first at
$r = r_{\rm p, 2}$ (either 1.2 or 1.05 AU)
and $\phi_{\rm p, 2} = \pi$ initially.

Our simulation grid spans the full $2\pi$
in azimuth, and extends from an outer radius
of 1.8 AU to an inner radius $r_{\rm in}$
that equals either 0.4 or 0.3 AU depending
on whether $r_{\rm p,1} = 1$ AU or 0.75 AU
(see \tabref{tab:setup}). 
Grid dimensions are 800 ($r$) $\times$ 3200 ($\phi$) when
$r_{\rm in}~=$ 0.4 AU,
and 960 $\times$ 3200 when $r_{\rm in}~=$ 0.3 AU.
Cells are spaced logarithmically in radius and
uniformly in azimuth.
Our choices yield a resolution of $\sim$18 cells
per scale height $h$
in both directions at $r = 1$ AU
(similar to the resolution of \citealt{Li09}).
Simulations at twice our standard resolution
did not produce
significant changes in either planet migration
or disk accretion rate for the first 100 yr.
We also tested our inviscid disk model without a planet,
and found that the numerical noise in $|\dot{M}|$ 
was about 3 orders of magnitude below planet-driven
disk accretion rates,
corresponding to a numerical viscosity of $\alpha<10^{-5}$.

Radial boundary conditions require special care in this study.
After experimenting with a few ways to measure disk accretion rates,
we found that the most stable method was to track the total
disk mass within a cylinder of radius 0.6 AU---a distance
intermediate between
the innermost planet and the inner disk boundary---while
preventing mass from leaving the grid.
We adopt ``zero flux'' boundary conditions where mass and momentum
fluxes across the inner and outer disk edges are always zero.
In \texttt{PEnGUIn}, this is achieved by solving a special 
Riemann problem at the boundaries, one where no wave
travels toward the simulation domain, and where the radial velocity
outside the domain is always zero. This implementation
conserved the total mass within the simulation domain
to numerical accuracy.
The accretion rate $\dot{M}$ at $r = 0.6$ AU
is calculated by following over time
the disk mass enclosed, $M_{\rm 0.6\, AU}(t)$.
Because the function $M_{\rm 0.6\,AU}(t)$ fluctuates strongly, we fit
independent lines to segments of data each lasting 20 yr,
taking $\dot{M}$ from the best-fitting slopes.

As a planet repels material away from its orbit, our boundary
conditions result in gas piling up at the inner and outer boundaries.
Our results can only be trusted to the extent that
these boundary pile-ups do not interfere with planet migration and disk
accretion. We therefore limit ourselves to studying only the
first few thousand years of planet-disk interactions, before 
boundary effects become too large.

\section{Results}
\label{sec:results}

We assess to what extent planets migrate in inviscid disks
(\S\ref{sec:migrate}),
and study how planet-driven accretion rates evolve with time and
depend on disk and planet masses (\S\ref{sec:accrete}).
For planet migration, models \#1--3 demonstrate differences
between viscous and inviscid disks, and between single-planet and two-planet
systems. For planet-driven accretion, we vary disk and planet masses
in models \#3--6 to test \eqnref{eqn:ana_mdot_wnum}.

\subsection{Planet migration}
\label{sec:migrate}

\begin{figure}[]
\includegraphics[width=0.99\columnwidth]{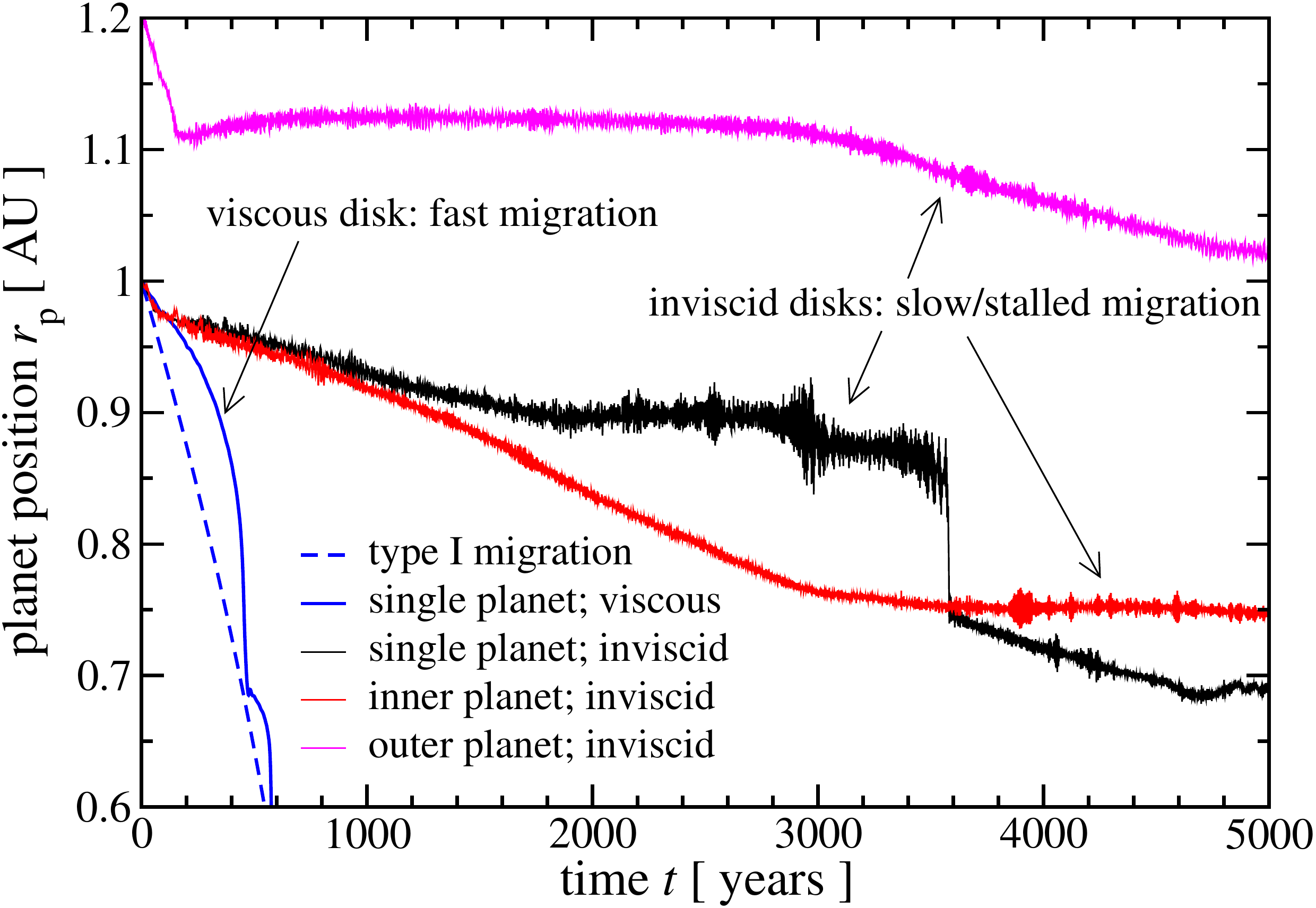}
\caption{Type I migration can be defeated for super-Earths
in inviscid disks. Shown here are the orbital radii
vs.~time of planets in three of our simulations:
a single $10 M_\oplus$ planet in a
gas-rich ($\Sigma_0 = 8.5 \times 10^3$ g cm$^{-2}$)
viscous ($\alpha = 10^{-3}$) disk (blue, model \#1);
a single planet of the same mass in a gas-rich
inviscid ($\alpha = 0$) disk
(black, model \#2); and two such planets in a gas-rich
inviscid disk (red+magenta, model \#3).
The blue dashed line is the theoretically expected
trajectory from Type I migration
(the integral of \eqnref{eqn:typeI} with $C=2$).}
\label{fig:mig}
\end{figure}

\begin{figure*}[]
\includegraphics[width=1.99\columnwidth]{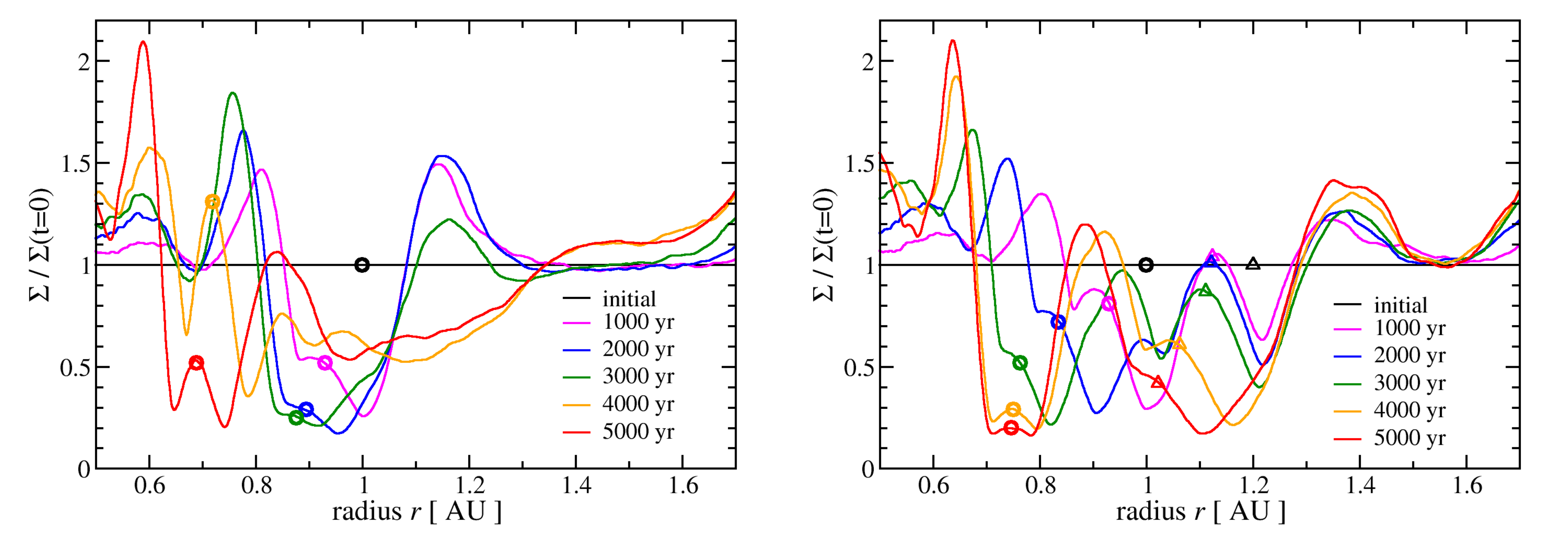}
\caption{Left: Surface density profiles, azimuthally averaged
and normalized against the initial power-law profile,
measured at 1000-yr intervals for our single-planet,
inviscid, gas-rich disk model (\#2). Right: Analogous profiles for
our two-planet, inviscid, gas-rich disk model (\#3).
Circles (triangles) mark the locations of the inner (outer)
planet.}
\label{fig:sigma}
\end{figure*}

\begin{figure}[]
\includegraphics[width=0.99\columnwidth]{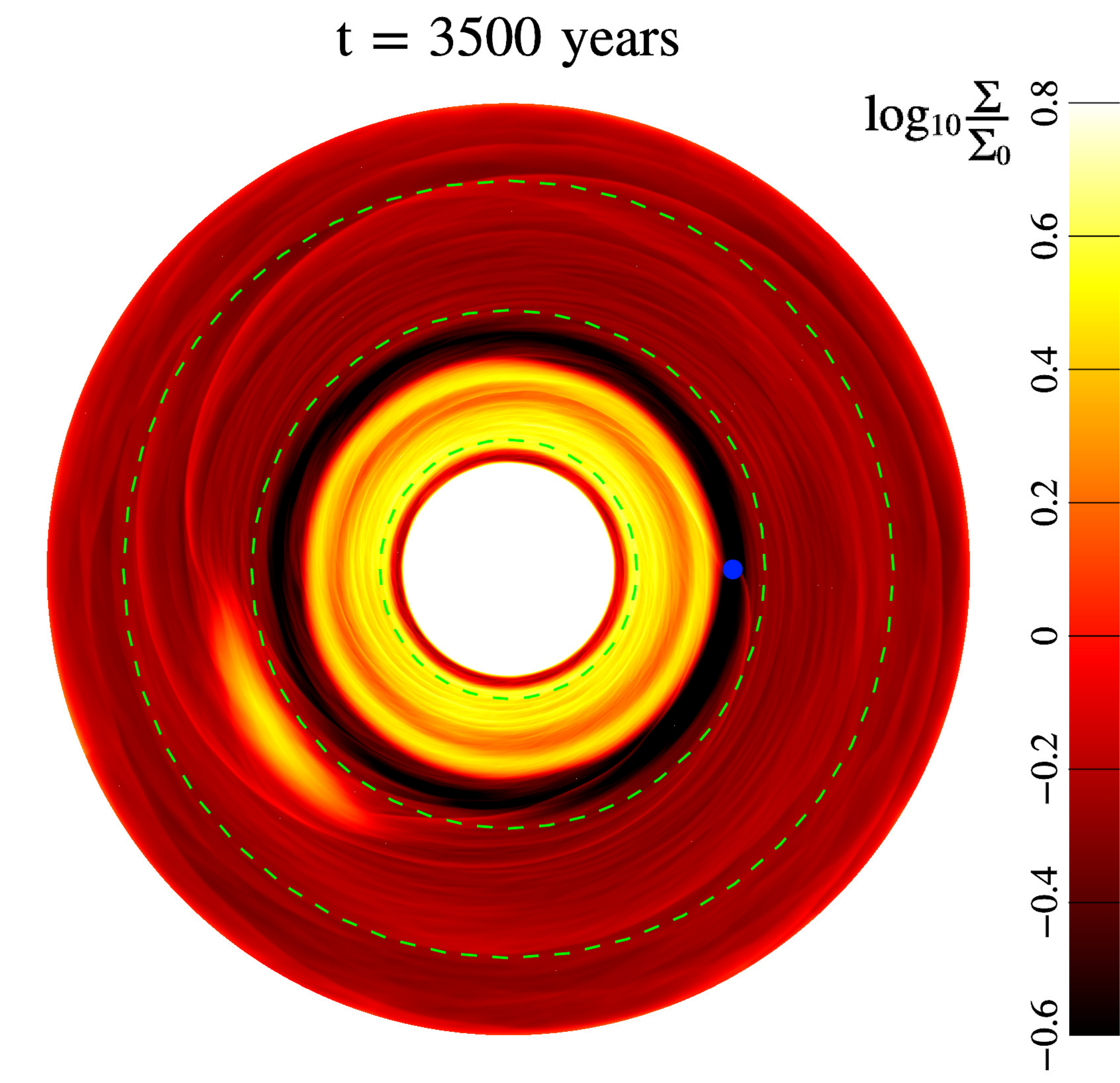}
\caption{Surface density snapshot of our single-planet, gas-rich simulation (model \#2) at 3500 years. Green dashed circles are at 0.5, 1, and 1.5 AU. The blue dot indicates the planet's position. This snapshot is taken near the time of the planet-vortex close encounter. The vortex, of approximate mass 100 $M_\oplus$, is seen at around 1.1 AU. After the encounter, the planet is perturbed radially inward (\figref{fig:mig}). The vortex ultimately decays away (\figref{fig:endpoint}).}
\label{fig:vortex}
\end{figure}

\begin{figure*}[]
\includegraphics[width=1.99\columnwidth]{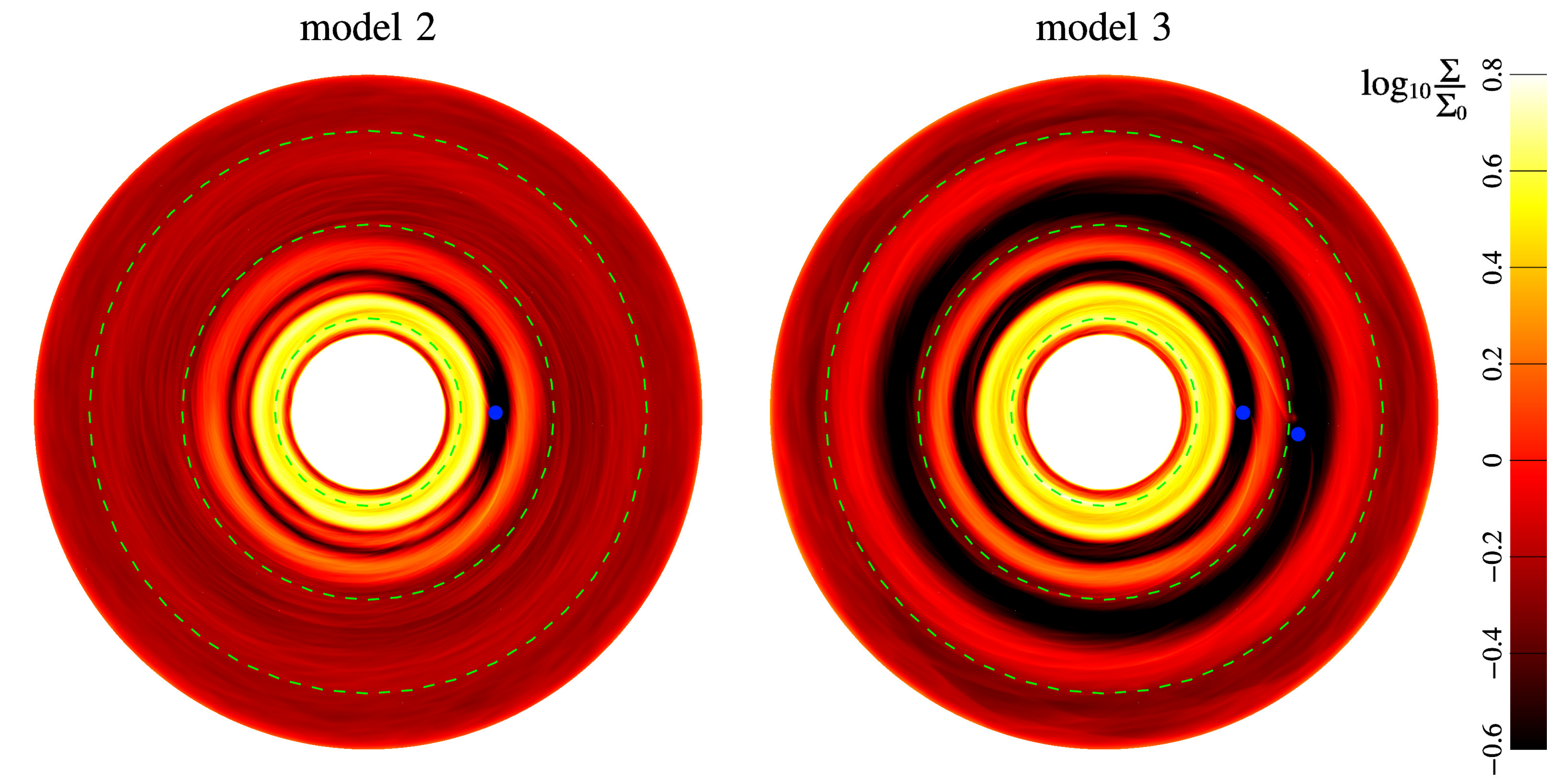}
\caption{Surface density snapshots of our inviscid gas-rich
simulations, for our single-planet model (\#2, left) and two-planet
model (\#3, right) cases, taken at their end time of 5000 years.
Green dashed circles are at 0.5, 1, and 1.5 AU. The blue dots indicate the planets' positions. In the single-planet snapshot, note that the vortex seen in \figref{fig:vortex} has completely decayed away. No comparably strong vortex was found in the two-planet simulation at any time, and the density field appears more axisymmetric than in the single-planet case.} 
\label{fig:endpoint}
\end{figure*}

\figref{fig:mig} plots the orbital evolution of 
planets, each of mass $10 M_\oplus$, 
in the gas-rich disk models (\#1--3). Overplotted
for comparison is the trajectory expected from integrating
the Type I migration rate,
\begin{equation}\label{eqn:typeI}
\dot{r}_{\rm p, Type \,I} = -2 C r_{\rm p} \Omega_{\rm p} \left(\frac{\Sigma_{\rm p} r_{\rm p}^2}{M_{\rm p}}\right) \left(\frac{M_{\rm p}}{M_*}\right)^2 \left(\frac{h_{\rm p}}{r_{\rm p}}\right)^{-2} \, ,
\end{equation}
using the unperturbed surface density law in
\eqnref{eqn:initial_sigma} to evaluate $\Sigma_{\rm p}$
(using the actual surface density in the viscous
disk simulation \#1 would give practically identical
results, since gaps do not form in that model).
Three-dimensional simulations suggest that
$C\sim 2$--3 for our given disk
profile \citep{DAngelo10,Fung17}; the blue
dashed curve in \figref{fig:mig} uses $C=2$.
\figref{fig:sigma} displays
azimuthally averaged surface density profiles at various
epochs in gas-rich, inviscid disk models \#2 and \#3,
with planet locations marked.

In agreement with the simulations of \citet{Li09}
and \citet{Yu10}, the planet
migration rate in our viscous disk (model \#1)
is similar to the Type I rate (punctuated by what appear
to be episodes of even faster Type III migration;
\citealt{Masset03,Peplinski08d}), and
much slower in inviscid disks (models \#2
and \#3). The initially rapid migration 
seen in the inviscid simulations at $t \lesssim 200$ yr
is a transient that decays after disk surface densities
adjust to planetary Lindblad torques, i.e., 
after the surface density pile-up ahead of the planet 
attains a fractional amplitude on the order of unity.
After this initial adjustment period,
migration slows and even stalls at times, with radial positions
changing by $\sim$10--30\%, or less, over kyr timescales.

In the single-planet, gas-rich simulation (black curve
in \figref{fig:mig},
model \#2), the planet journeys slowly inward for the first
$\sim$2000 yr and practically stops in the mean
from $t \simeq 2000$--3500 yr,
as disk gas that the planet has pushed inward to
$r \simeq 0.7$--0.8 AU piles up (blue and
green curves in the left panel of \figref{fig:sigma})
and stymies further migration.
At $t \simeq 3500$ yr, the planet experiences a sudden
drop in orbital radius; we traced this drop to
a close encounter between the planet and a vortex formed at its outer gap
edge at $r \simeq 1.1$ AU (see \figref{fig:vortex}).
Some time after the encounter, the vortex gradually disperses,
completely decaying away by the end of our
simulation at 5000 yr (\figref{fig:endpoint}, left panel).
Similar planet-vortex interactions were found
in simulations by \citet{Lin10} and \citet{Yu10}.
Thereafter, the planet's migration returns to its
near-zero mean pace.

In our two-planet, gas-rich simulation (\#3),
the outer planet stalls for the first $\sim$3000 yr (magenta
curve in \figref{fig:mig}), apparently trapped
at a local surface density maximum created
by the inner planet whose forcing 
dominates: see the magenta curve
in the right panel of \figref{fig:sigma},
and note how similar it is to the corresponding
magenta curve in the left panel for the single-planet case.
Gas pushed inward by the outer planet strengthens the
torque on the inner planet
and forces the latter
to migrate inward by $\sim$20\% over the same time period
(red curve in \figref{fig:mig});
contrast this behavior with the stalling observed
in the single-planet case (black curve). 
Eventually, at $t \sim 4000$ yr,
the outer planet disperses the surface density
maximum in its vicinity, and proceeds to migrate
slowly inward, slowing down near $t\sim 5000$ yr
as it runs into sandwiched gas. Meanwhile,
the inner planet ultimately comes to a near halt
in much the same way that it does in the single-planet 
simulation, having run into material that has piled up
just interior to its orbit.
The right panel of \figref{fig:endpoint}
shows the final surface density distribution.
We emphasize that this pile-up is physical as it
is located at $r \simeq 0.6$ AU, away
from the inner grid boundary of the simulation at
$r_{\rm in} = 0.4$ AU.
The latter location has
its own separate pile-up, which does not grow
to significance over the limited duration of our simulations.
The same statement applies to the outer grid boundary.

We note that in none of the two-planet simulations did we observe
the formation of a vortex like the one seen in our single-planet
simulation. This difference might be physical,
and deserves attention in future studies of planet-vortex 
interactions.

All other factors being equal, lower disk masses should lead
to even slower planetary migration rates. 
This is confirmed in our gas-poor simulations (models \#4--6)
which exhibit no measurable change in planet mean
radial positions. Thus the gas-poor simulations
can be used to diagnose disk accretion rates
without the complicating effects of planetary migration,
as we discuss in the next subsection.

\subsection{Disk accretion}
\label{sec:accrete}

\begin{figure*}[]
\includegraphics[width=1.99\columnwidth]{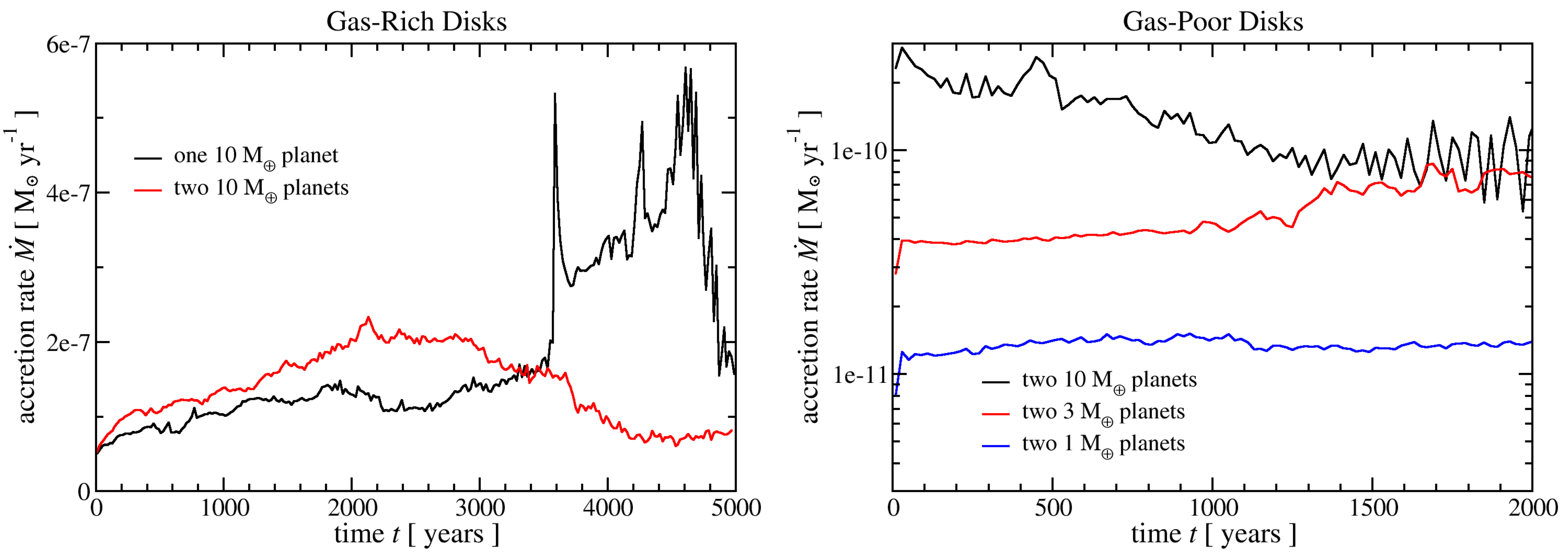}
\caption{Disk accretion rates measured at $r = 0.6$ AU,
a location interior to all planets at all times.
Each data point represents an average
over 20 years (see \S\ref{sec:initial}).
Left: Gas-rich $\alpha=0$ simulations (model \#2 in black, \#3 in red).
Right: Gas-poor $\alpha=0$ models (\#4--6 in black, red, and blue,
respectively). At $t \lesssim 1000$ yr, accretion rates
agree with predictions from \eqnref{eqn:ana_mdot_wnum} to within
a factor of 2. Variations in $\dot{M}$ at later times
reflect planet migration (compare with \figref{fig:mig})
and deepening of gaps.}
\label{fig:acc}
\end{figure*}

\begin{figure*}[]
\includegraphics[width=1.99\columnwidth]{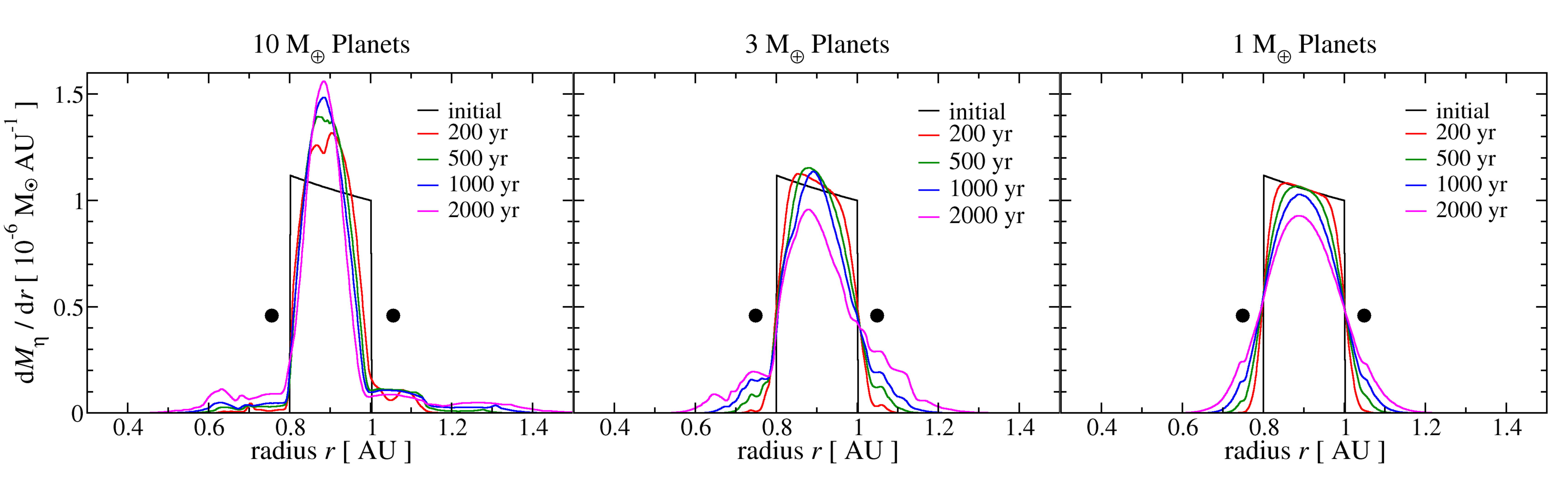}
\caption{Tracking the material initially sandwiched by two planets. From
left to right, we show results from models \#4, 5, and 6. Black circles
denote the planets' locations which do not
change in these gas-poor runs. In all cases,
the gas escapes over time to either side of the planetary pair.}
\label{fig:eta}
\end{figure*}

\figref{fig:acc} shows disk accretion rates
as functions of time for all our inviscid models. 
As described in \S\ref{sec:initial}, the accretion
rate $\dot{M}$ is measured by tracking the build up
of disk mass inside $r = 0.6$ AU, a location
interior to the planets at all times. 
Later, in \S\ref{sec:sandwich},
we track the movement of mass initially
between two planets.

We begin by checking whether
our simulations are compatible with the analytic expectation
for $\dot{M}$ given by \eqnref{eqn:ana_mdot_wnum}.
The comparison is best made at early times of the simulation,
$t \lesssim 1000$ yr,
before radial surface density profiles
become too distorted.
For the single-planet, gas-rich
model \#2, \eqnref{eqn:ana_mdot_wnum} yields
$\dot{M} \sim 3 \times 10^{-8} \msunyr$. This
prediction is within a factor of $\sim$2 of the 
simulated result at early times
(left panel of \figref{fig:acc}).
As for the corresponding two-planet model \#3,
we expect from (\ref{eqn:ana_mdot_wnum})
that $\dot{M}$ should be only fractionally
larger than for the single-planet case; the
second planet is farther removed from where
we measure $\dot{M}$ ($r = 0.6$ AU),
and so makes only a $\sim$50\%
contribution to the accretion flow there as
compared to the inner planet.
This is approximately consistent with \figref{fig:acc}.

For our gas-poor model \#4, \eqnref{eqn:ana_mdot_wnum} predicts
$\dot{M} \sim 1.5 \times 10^{-10} \msunyr$, again
within a factor of 2 of the simulated result (right panel of
\figref{fig:acc}). Scaling the planet mass $M_{\rm p}$
down by a factor
of 10 from models \#4 to \#6 should, according to
\eqnref{eqn:ana_mdot_wnum}, reduce $\dot{M}$ by a factor of
$10^{3/2} \simeq 30$. By comparison, \figref{fig:acc}
shows a factor of $\sim$20
decrease between these two models; we consider
this acceptable agreement with the analytic expectation.
In summary, our simulations
support the various functional dependencies
predicted by \eqnref{eqn:ana_mdot_wnum}
to within a factor of 2.

At later times, $t \gtrsim 1000$ yr, 
we observe time variability in $\dot{M}$ caused by
the deepening of planetary gaps, and by
planet migration. These variations are limited
to factors of a few. 
The simulations easiest to interpret are models \#4--6 (right
panel of Figure \ref{fig:acc}) which have too little disk gas
to drive planet migration. The initial gradual decline in $\dot{M}$
seen in model \#4 is caused by the deepening
of gaps opened by its 10-$M_\oplus$ planets;
over the course of 2000 yr, the gas density in the immediate vicinity
of the planets decreases by a factor of $\sim$5 for the inner planet 
and by a factor of $\sim$3 for the outer one.
Models \#5 and \#6 exhibit steadier accretion rates,
as their planets have lower masses which are less effective
at opening gaps.

More complicated behavior is seen in the gas-rich simulations
where planets migrate more appreciably.
Comparison of Figures \ref{fig:mig} and \ref{fig:acc}
reveals that increases in $\dot{M}$ can be traced to planets
moving inward, either gradually, as in the first 2000 yr
of models \#2 and \#3, or suddenly, as in the
planet-vortex encounter at $t \simeq 3300$ yr in model \#2.
Decreases in $\dot{M}$ correspond to planets opening gaps
upon moving to new locations.

\subsubsection{The fate of gas initially sandwiched between planets}
\label{sec:sandwich}
Disk accretion driven by planets would be impractical
if material residing between planets were unable to escape.
We track this sandwiched gas in simulations
\#4--6, each containing a pair of planets which migrate
negligibly. We assign each gas parcel a ``passive scalar'' $\eta$
that equals 1 for gas initially located between 
$r = 0.8$ and 1 AU (between the two planets), and is $0$ everywhere else.
Gas elements carry $\eta$ as a conservative quantity.

\figref{fig:eta} follows the $\eta$-tagged gas by plotting 
\begin{equation}
\label{eqn:m_eta}
\frac{\partial{M_\eta}}{\partial r} = \int^{2\pi}_{0}\eta \Sigma r~d\phi \, ,
\end{equation}
vs.~$r$ at various times.
We find that the sandwiched gas 
is torqued both inward and outward,
escaping in roughly equal amounts
to the inside of the inner planet
and to the outside of the outer planet.
The opposing torques from the
two planets do not in general cancel.
The opposing torques from the  
two planets do not in general cancel, although 
in model \#4 containing the highest mass planets,
some gas does concentrate along the midline between
the planets in a ``shepherded'' ring, resulting in less
mass leaking out of the sandwiched region. 
Looking at models \#4--6 in Figure \ref{fig:eta},
we see no clear trend between the rate at which sandwiched gas escapes
and planet mass. There seems to be a complicated confluence of effects
in the sandwiched region. Lindblad torques act to shepherd some gas
while also opening gaps that reduce the local gas density; and
co-orbital torques allow gas to escape via horseshoe orbits,
whose libration times scale only weakly with planet mass  
($t_{\rm lib} \propto M_{\rm p}^{-1/2}$; \citealt{Paardekooper09a}).
A detailed analysis is deferred to another paper; for now,
we conclude that, at least for comparable mass planets
with orbital spacings like the one we have assumed,
the inner planet presents a porous barrier
to material pushed inward by the outer planet.
Apparently gas that is pushed by the outer planet
toward the inner planet can be shuttled
past the latter on horseshoe orbits (and vice versa).

\vspace{0.2in}

\section{Summary and Discussion}
\label{sec:conclude}

Using hydrodynamical simulations, we have demonstrated
that super-Earths in inviscid disks can simultaneously avoid type I migration
(\figref{fig:mig}) and promote disk accretion (\figref{fig:acc})
by driving density waves. Disk accretion rates
measured from our simulations verify analytic predictions
(\eqnref{eqn:ana_mdot_wnum}) to within a factor of 2 .
We observed gap opening and planet migration in inviscid disks
to be modest and to introduce order-unity effects on the disk
accretion rate. We also found in our two-planet simulations
that material initially sandwiched between two planets leaks past both
into the innermost and outermost disks (\figref{fig:eta}).

Our models omit a number of effects. Many of these are not overly
concerning. Although our simulations are 2D, no substantive difference
between 2D and 3D treatments of planet-disk interactions 
in viscous disks has been reported
vis-\`{a}-vis gap opening \citep{Fung16a}
or planetary torques \citep{Fung17}. Our neglect of disk self-gravity
should be an excellent approximation, as
the Toomre $Q$-values of our disks greatly exceed unity.
Our planets are not allowed to accrete gas,
but super-Earths/sub-Neptunes
are inferred observationally to have only modest amounts
of gas---less than 10\% by mass---acquired gradually
over the entire disk lifetime \citep{Lee16}.

More interesting frontiers to pursue include
incorporating disk thermodynamics, 
as radiative cooling and differential heating across gap walls
are thought to materially affect planet-disk interactions
(e.g., \citealt{Kley12}; \citealt{Tsang14}).
Of course, extending the durations of the simulations,
and including more planets with different orbital
architectures, would also be welcome, for greater realism and
to enable more direct connections with observations.
Closer study of gap depths is warranted;
we observed surface density contrasts only on the order
of unity (\figref{fig:sigma}), in clear deviation from scaling relations
derived from viscous disks \cite[e.g.,][]{Fung14},
and surprising insofar as less viscous gas should be
less effective at diffusively back-filling gaps.
The shallowness of the gaps is due partly to the planets migrating 
and re-starting the gap-opening process at each new radial location \citep{Malik15}.
But how much of it is due to the limited duration of our
simulations ($\leq 5000$ yr), or to hydrodynamical instabilities
like the Rayleigh instability \citep{Fung16a}, remains
to be worked out.
Finally, survival against planetary migration
is not guaranteed: planets with mass $M_{\rm p} > M_{\rm cr}$
(\eqnref{eqn:m_cr}) stall but less massive planets do not.
The question is whether rocky planets can
coagulate fast enough to cross the $M_{\rm cr}$ threshold
before they succumb to migration.

Our results support the proposal
by \citet{Goodman01} that predominantly rocky
planets---super-Earths and Earths---can solve, or at least help
to solve, the problem of how protoplanetary gas disks ultimately disperse.
Given that a single planet can push gas over a lengthscale of
approximately half its orbital radius,
shuttling gas from 5 AU down to 0.1 AU would require
about 6 super-Earths distributed in roughly equal logarithmic intervals
across this distance. Such planet multiplicities
are reasonable, given the profusion of super-Earths/sub-Neptunes
discovered by {\it Kepler} \citep[e.g.,][]{Pu15}.
To be sure, a disk accretion flow driven by a planetary system
will be unsteady, changing not only on secular, Myr-long timescales,
but also on much shorter ones, with mass alternately
accumulating and dispersing
in interplanetary space, as we have seen in our simulations
(\figref{fig:sigma}).
No matter how complicated the history, however,
all gas must ultimately be torqued out of
sufficiently compact planetary systems.
It may be torqued by planets
so far inward that turbulence driven by the
magneto-rotational instability, activated in the innermost
regions which are sufficiently thermally ionized
\citep[e.g.,][]{Desch15},
takes over the job of disk accretion onto the host star.
Or it may be torqued by planets
so far outward that it escapes from the system
altogether in a photoionized wind \citep[e.g.,][]{Alexander14}.

How massive a gas disk can a set of super-Earths drain?
A first consideration is that the disk surface density
can not be so large that the embedded planet
mass $M_{\rm p} < M_{\rm cr} \propto \Sigma_{\rm p}^{5/13}$,
lest the planet migrate away. Based on the typical
parameters listed in \eqnref{eqn:m_cr}, a disk
containing $\Sigma_{\rm p}r_{\rm p}^2 \sim 10^{-3} M_\ast \sim 300 M_\oplus$
of gas could be evacuated by $\sim$6 super-Earths
weighing a total of $\sim$$30 M_\oplus$.
In such an initially gas-rich environment, we anticipate
the planets would migrate to and fro by a few tens of percent
in orbital distance (\figref{fig:mig}).
As the disk drains in the long term, whatever
slow and erratic migration the planets
undergo diminishes.
Most of the angular momentum
of interplanetary gas would be transported
to the outermost disk, exterior to all the planets, 
either by Lindblad torques or by direct advection.

The large cavities of transitional disks may have been
excavated over time by families of super-Earths. 
Observationally, gas densities inside cavities
can be suppressed relative to their values
outside by two to four orders of magnitude
(e.g., \citealt{Carmona17}; \citealt{Dong17c}).
To reproduce these strong depletions,
appeal is commonly made to giant planets in viscous 
$\alpha$-disks that can open deep gaps \citep[e.g.,][]{Dong16,Dong17b}.
But an alternative interpretation is that the cavities
have been eroded gradually over time by much smaller mass planets
in inviscid disks. So far as we have measured in our simulations,
such planets open gaps having only order-unity
surface density contrasts. Nevertheless, given
sufficient time, they can drain interplanetary gas
by orders of magnitude.
Because planet-driven accretion rates $\dot{M}$ scale linearly
with gas surface densities $\Sigma$, we have $\dot{\Sigma} \propto -\Sigma$
which implies exponential decay of the gas content.
If it takes $t_{\rm e-fold} \sim 10^5$ yr
to reduce a total disk mass of $10^{-3} M_{\odot}$ by a factor of $e$
(this assumes a contemporaneous mass transport rate
of $10^{-8} \msunyr$),\footnote{This transport can be inward
or outward---it should not matter as long as the region occupied
by the planets is monotonically drained of gas over time.}
then it takes $7 t_{\rm e-fold} \sim 7 \times 10^5$ yr to reduce
it by a factor of 1000.
Another feature of this picture is that if 
super-Earths drain disk mass faster than it takes for their
nascent atmospheres to
cool and acquire more mass,
they may be able to 
forestall
runaway accretion \citep{Lee14}.
Depleting the local disk density by a factor of $\gtrsim 100$
(relative to the minimum-mass extrasolar nebula) 
over timescales of $\sim$1 Myr suffices to keep
the gas-to-solids mass fraction of super-Earths
$\lesssim 10\%$, in accord with observations \citep{Lee16}.

One potential problem with this scenario
is that it predicts mass accretion rates to lower in proportion
to disk gas densities. Although some transitional
disks do have low accretion rates \citep[e.g.,][]{Dong17c},
others do not, with a few having $\dot{M}$ as high
as $10^{-7} \msunyr$ (e.g., \citealt{Rosenfeld14}; \citealt{Carmona17};
\citealt{Wang17}). Even so, system-to-system
variations in orbital architectures,
particularly in the masses of the orbiting
companions, could help to resolve
this problem. We have focused here on super-Earths because
they are commonplace, but in principle gas giants and perhaps brown
dwarfs or even low-mass stars may serve in their stead.\footnote{The 
case of the transitional disk HD 142527 is especially intriguing:
its host star accretes at a rate of $\dot{M} \sim 10^{-7} \msunyr$
\citep{GarciaLopez06},
and its cavity contains a $0.2 M_\odot$ companion
highly inclined to the disk \citep{Casassus15,Lacour16}.}

That disk gas is intrinsically inviscid (laminar)
is suggested on other, independent grounds.
Large-scale asymmetries in transitional disks (e.g.,
\citealt{Casassus13}; \citealt{vanderMarel13}; \citealt{Pinilla15})
have been interpreted as vortices (e.g., \citealt{Zhu16};
\citealt{Baruteau16}), but these vortices
are spawned only in low-viscosity disks. 
\citet{Zhu16} found that the Shakura-Sunyaev viscosity
parameter $\alpha$ needed to be $10^{-4}$ or lower before vortices could
grow from sharp density gradients.
In other news, attempts to detect turbulence in the outer portions
of disks using molecular line observations 
have so far come up empty-handed (\citealt{Flaherty15};
Flaherty et al.~2017, in preparation).
And \citet{Rafikov17}, in a systematic analysis of disk
accretion rates and masses, suggests that accretion
may not proceed viscously (i.e., diffusively),
but may be enabled instead by 
spiral density waves and/or disk winds.
All these recent developments, in addition to our present work, suggest 
that the reason the community has not discovered
a robust explanation for a non-zero $\alpha$ in protoplanetary
disks is that none exists: that in fact such disks are for the most part
inviscid, and accrete 
primarily by the action of gravitational torques,
exerted either by disk gas itself at early times
(e.g., \citealt{Gammie01}; \citealt{Cossins09}),\footnote{In 
self-gravitating disks, characterizing transport in terms
of a non-zero $\alpha$ is commonly done, but mostly
for convenience. Gravity is a long-range force and
not naturally captured within a local theory
like the one defining $\alpha$.}
or by planets at late times.

\acknowledgments 
We thank Pawel Artymowicz, Ruobing Dong, Eve Lee, Hui Li, Zhi-Yun Li, Fr\'{e}d\'{e}ric Masset, Norm Murray, Ruth Murray-Clay, Sijme-Jan Paardekooper, Roman Rafikov, Yanqin Wu, Zhaohuan Zhu, and an anonymous referee for encouraging discussions and helpful feedback.
This work was performed under contract with the Jet Propulsion
Laboratory (JPL) funded by NASA through the Sagan Fellowship Program
executed by the NASA Exoplanet Science Institute.
EC is grateful for financial support from NASA and NSF.

\bibliographystyle{apj}
\bibliography{Lit}

\end{document}